\documentclass{iopjournal}

\usepackage[T1]{fontenc}
\usepackage[utf8]{inputenc}
\usepackage{amsmath,amssymb}
\usepackage{booktabs}
\usepackage{subcaption}
\usepackage{float}
\usepackage[numbers,sort&compress]{natbib}
\usepackage{microtype}

\begin{document}

\articletype{Paper} 

\title{A full-wave ICRH solver with full kinetic accuracy using only first order differential equations}

\author{Liuheng Wu$^{1,*}$, Ziyu Wang$^{1,*}$, Huasheng Xie$^{2,\dagger}$ and Chang Liu$^{1,\dagger}$}

\affil{$^1$State Key Laboratory of Nuclear Physics and Technology, School of Physics, Peking University, Beijing 100871, China}

\affil{$^2$Beijing VeloAlpha Technology Co., Ltd., Beijing, 100080, China}

\affil{$^*$These authors contributed equally to this work.}

\affil{$^\dagger$Corresponding authors.}

\email{huashengxie@gmail.com; changliu@pku.edu.cn}


\begin{abstract}
Accurate full-wave computation of ion cyclotron
resonance heating (ICRH) typically requires solving
the computationally demanding integro-differential
dielectric tensor of the hot plasma. Based on
the method in [Phys. Plasmas 32, 082110 (2025)],
we develop BO-FW1D, a 1D frequency domain
finite difference full-wave solver which generates
full kinetic, accurate and efficient simulations
utilizing a fluid plasma model.
To accurately convert the conductivity tensor
in phase space into a first-order differential
equation, we introduce a mathematically rigorous framework
using pseudodifferential operator theory based on Kohn-Nirenberg
ordering. Benchmarks with an all-orders 1D full-wave solver
show high consistency in critical kinetic effects and
electric-field distributions, while substantial
computational acceleration is achieved. This work could
lay the foundation for developing full-wave
solvers that combine full kinetic accuracy with high
computational efficiency.
\end{abstract}

\section{Introduction}
\label{sec:intro}

Ion cyclotron resonance heating (ICRH) is one of the
core methods for heating, sustaining, and
controlling high-temperature plasmas, and accurately
simulating the propagation and absorption of
radio-frequency (RF) waves in the plasma is a key
component of ICRH modeling. To include critical
kinetic effects such as Landau damping~\cite{landau1946vibrations},
Bernstein modes~\cite{bernstein1958waves}, and mode
conversion process between fast waves and ion
Bernstein waves (IBWs)~\cite{brambilla1989finite,brambilla1999numerical,jaeger2001all},
full-wave solvers must evaluate
the plasma dielectric tensor incorporating the full
kinetic response, which is computationally expensive.

Traditional spectral full-wave codes, such as
AORSA~\cite{jaeger2001all}, describe the kinetic
dielectric tensor as an infinite series of truncated
modified Bessel functions of the first kind $I_n$.
Besides its high computational cost,
when treating high-$k_\perp$ IBWs,
the parameter $\lambda = k_\perp^2 \rho^2$
becomes large, the function $\Gamma_n(\lambda)
= e^{-\lambda} I_n(\lambda)$ decays rapidly in numerical
evaluation, making the accuracy difficult
to guarantee. TORIC expands the integro-differential
plasma current function by using the small Larmor radius
approximation, and retains the main finite Larmor radius
(FLR) effects through the Swanson-Colestock-Kashuba
(SCK) approximation~\cite{brambilla1989finite,brambilla1999numerical}.
However, this method is valid
only for $k_\perp \ll 1$, so it
underestimates the damping of short-wavelength
IBWs during propagation. Dirk Van Eester and E.~A.~Lerche
approximated the integro-differential kinetic
dielectric tensor using high-order polynomials
of $k_\perp$, retaining a higher-order differential
equation~\cite{van2021solving,bude2021accelerating}.
While retaining the full FLR response, the
fitting method is only valid for $k_\perp$ inside
the fitting range. On the other hand, although conventional
multi-fluid~\cite{xie2024plasma} and gyro-Landau fluid
(GLF)~\cite{hammett1990fluid} models are computationally
efficient, they cannot retain the complete FLR
effects and thus miss key kinetic wave modes such as
IBWs. Self-consistently incorporating kinetic effects
while avoiding explicit kinetic integrals constitutes a
long-standing core challenge in the field of ICRH
full-wave simulation.

To break this impasse, Xie recently proposed a novel
fluid-form treatment of the plasma dielectric
tensor~\cite{xie2025developing}. This method applies
Pad\'e approximations and pole expansions to
rationalize the computationally expensive functions,
effectively circumventing the numerical instability
while preserving an algebraic rational-fraction form.
A related approach has recently been pursued by
Shiraiwa \textit{et al.}~\cite{shiraiwa2026full},
who approximate the non-local dielectric response by
rational functions of $k_\perp^2$. However,
such an expansion is intrinsically symmetric in
$k_\perp$ and can therefore represent only even
responses. In Xie's model, the pole expansion
is first order in $k_\perp$, so that the
rational form carries no such symmetry constraint and
contains the full response. The ``fluid-like''
model incorporates Landau damping and finite Larmor
radius effects with only rational functions based on
frequency $\omega$, parallel wavenumber $k_\parallel$
and perpendicular wavenumber $k_\perp$. Meanwhile,
it can accurately attain the wave branches of
IBWs and electron Bernstein waves
(EBWs)~\cite{xie2025developing} when solving the
perpendicular wavenumber. Based on its fluid-like
form and its ability to accurately capture Bernstein waves,
the model is referred to as the ``fluid BW model''.

Nevertheless, this model has so far been validated only
for local, linear, electromagnetic
kinetic plasma waves in an infinite, homogeneous
system~\cite{xie2025developing}. Owing to its ability
to incorporate accurate kinetic effects while
remaining valid for arbitrary $k_\perp$, the fluid BW dielectric
tensor is expected to enable faster and more accurate
full-wave simulation for ICRH.
Extending this fluid BW dielectric tensor to
inhomogeneous full-wave simulation
constitutes the starting point of the present work.
The compact rational-fraction algebraic structure of
the fluid BW model makes it possible to convert
the local dispersion relation into a first-order
local differential equation in real space, turning
the system into a sparse banded matrix that can be
solved at a cost linear in the number of grid points.

The straightforward route is to treat the wavenumber
as a differential operator rather than a number,
$k_x \rightarrow -i\partial_x$, and to rewrite the
local dispersion relation as a differential equation
in real space. However, in an inhomogeneous plasma the
coefficients depend on $x$ and do not commute with
$-i\partial_x$, so that operator compositions cannot
be treated as ordinary algebraic products. If so,
a non-physical amplification of the IBW amplitude will
occur. To this end, we introduce the Kohn-Nirenberg
pseudodifferential operator theory~\cite{folland1989harmonic},
which provides a rigorous mathematical framework for
the conversion from phase-space symbols to real-space
operators. The correction term due to operator
non-commutativity is derived under this framework,
and the non-physical amplification is eliminated.
As a result, the finite-difference solver built on the
corrected equations, BO-FW1D, achieves high consistency
with the all-orders benchmark code under the JET
parameters, demonstrating
both the necessity and the effectiveness of the
conversion framework.

The remainder of this paper is organized as follows.
Section~\ref{sec:model} presents the fluid BW dielectric tensor and
derives the corrected response
equation in real space. Section~\ref{sec:results} benchmarks the
resulting solver in two steps. First, the overall agreement
with the all-orders benchmark code together with
the necessity of the operator composition correction
term. Second, the computational scaling and grid
convergence of the solver.
Section~\ref{sec:conclusion} summarizes the main findings and discusses
the limitations of the present one-dimensional
treatment together with possible extensions and future works.

\section{Model and Equations}
\label{sec:model}

\subsection{The Fluid BW Model}

The core of fluid BW model~\cite{xie2025developing} is to
apply rational approximations to the plasma
dispersion function $Z(\zeta)$ and to $\Gamma_n(\lambda)
= e^{-\lambda} I_n(\lambda)$, thereby
transforming the kinetic dielectric tensor from its
integral form into an algebraic fractional form.
The plasma dispersion function $Z(\zeta)$ is expanded
into a sum of partial fractions via a J-pole
approximation, whose coefficients $b_j$ and $c_j$
have been derived in Ref.~\cite{xie2024rapid}:
\begin{equation}
Z(\zeta) \simeq \sum_{j=1}^{J} \frac{b_j}{\zeta - c_j}
\label{eq:zpoles}
\end{equation}

The function $\Gamma_n(z^2)$ is expressed via an
$L$-pole expansion ($n \geq 0$, with $\Gamma_{-n} =
\Gamma_n$):
\begin{equation}
\Gamma_n(z^2) = \sum_{l_n=1}^{2n+L} \frac{r_{l_n}}{z -
p_{l_n}}
\label{eq:gammapoles}
\end{equation}

where $z = k_\perp |\rho_{cs}|$, $\rho_{cs}$ is the
Larmor radius of species $s$. The poles
$p_{l_{n}}$ and residues $r_{l_{n}}$ are fixed by the
rational-expansion procedure of Ref.~\cite{xie2025developing}.
Under these rational approximations, the full kinetic
conductivity tensor in a uniform, Maxwellian
plasma is rewritten as a sum of rational fractions:
\begin{equation}
\boldsymbol{\sigma} = \epsilon_0 \sum_l
\begin{bmatrix}
\dfrac{b_{lxx}}{k_x - c_l} & \dfrac{b_{lxy}}{k_x - c_l}
& \dfrac{b_{lxz}}{k_x - c_l} \\[8pt]
\dfrac{b_{lyx}}{k_x - c_l} & \dfrac{b_{lyy}}{k_x - c_l}
& \dfrac{b_{lyz}}{k_x - c_l} \\[8pt]
\dfrac{b_{lzx}}{k_x - c_l} & \dfrac{b_{lzy}}{k_x - c_l}
& \dfrac{b_{lzz}}{k_x - c_l}
\end{bmatrix}
\label{eq:conductivity}
\end{equation}

where the summation index $l$ runs over the particle
species $s$, the truncated Bessel harmonic
order $n = 0, 1, \dots, N$, and the partial-fraction
terms $l_n$ of each $\Gamma_n$:
\begin{equation}
\sum_l \equiv \sum_s \sum_{n=0}^{N} \sum_{l_n}
\label{eq:sumindex}
\end{equation}

The coefficients $b_{l\alpha\beta}$ and $c_l$ are fully
determined by the plasma parameters
(density, temperature, and magnetic field) and the
expansion coefficients. Their complete expressions
are given in Ref.~\cite{xie2025developing}.

Expanding the perturbed plasma current with the same
index $l$ as $J_\alpha = \sum_l v_{l \alpha}$
($\alpha = x, y, z$) and using the conductivity relation
$\boldsymbol{J} = \boldsymbol{\sigma} \cdot
\boldsymbol{E}$, the algebraic relation in
rational-fraction form for each level $l$
can be constructed. For notational simplicity, the vacuum
permittivity $\epsilon_0$ is hereafter
absorbed into the coefficients
$\overset{\leftrightarrow}{b}_l$:
\begin{equation}
\boldsymbol{v}_l =
\frac{\overset{\leftrightarrow}{b}_l}{k_x - c_l} \cdot
\boldsymbol{E}
\label{eq:fluid-unit}
\end{equation}

Equation~\eqref{eq:fluid-unit} constitutes the basic unit of the
fluid BW dielectric tensor. In Xie's work,
this equation, together with Maxwell's equations, is
cast into the matrix eigenvalue problem $
\boldsymbol{M} \cdot \boldsymbol{X}=k_x
\boldsymbol{X} $,
which accurately calculates the dispersion
relations of IBWs and EBWs in uniform media.

\subsection{Differential Response Equation}

To exploit the algebraic structure of Eq.~\eqref{eq:fluid-unit}
for solving the wave field in
inhomogeneous media, the wavenumber $k_x$ must be
replaced by the differential operator
$-i\partial_x$, while the coefficients
$b_{l\alpha\beta}$ and $c_l$ become functions of $x$.
If the denominator in Eq.~\eqref{eq:fluid-unit}
is first cleared algebraically before making
this substitution, one obtains the first-order
differential equation:
\begin{equation}
(-i\partial_x - c_l(x)) \boldsymbol{v}_l(x) =
\overset{\leftrightarrow}{b}_l(x) \cdot
\boldsymbol{E}(x)
\label{eq:naive-clearing}
\end{equation}

However, when Eq.~\eqref{eq:naive-clearing} is directly discretized
with a central-difference scheme and solved,
the computed IBW amplitude exhibits a
significant non-physical amplification, which will be
demonstrated in Section~\ref{sec:results}. The root of this
problem lies in the fact that, in the $k_x$
matrix eigenvalue problem, $k_x$ in Eq.~\eqref{eq:fluid-unit} is
an algebraic variable, so that $(k_x - c_l)$
commutes freely with the other coefficients and the
clearing operation is equivalent to simple
algebraic multiplication and division. In the full-wave
problem with inhomogeneous media, however,
$-i\partial_x$ is a differentiation rather than an
algebraic multiplication: it acts on every
$x$-dependent quantity to its right, including the
field $E(x)$, the current $\boldsymbol{v}_l(x)$,
and the coefficients $b_{l\alpha\beta}(x)$.
The algebraic simplification rules
that hold in uniform media lose their validity once
differentiation is introduced, because the order
of multiplication can no longer be freely interchanged.
Hence, the local rational-fraction
dispersion relation cannot be extended to an
inhomogeneous differential equation by simple algebraic
means.

It is therefore necessary to introduce an appropriate
mathematical framework. To this end, we adopt the
pseudodifferential operator theory based on
Kohn-Nirenberg (KN) ordering\cite{folland1989harmonic}.
The reason why we choose
this framework is that it links an arbitrary phase
space function $\sigma(x, k)$ with a real-space
operator $\sigma(X, D)$ through the following equation:
\begin{equation}
\sigma(X, D) f(x) = \int_{-\infty}^{+\infty} e^{ikx}
\sigma(x, k) \hat{f}(k) \, dk
\label{eq:kn-quantization}
\end{equation}
where $f(x)$ is an arbitrary function and $\hat{f}(k)$
denotes its Fourier transform. The capital letters $X$
and $D$ denote the coordinate and differentiation
operators, respectively,
\begin{equation*}
X f(x) = x f(x), \qquad D f(x) = -i \partial_x f(x),
\end{equation*}
so that $\sigma(X, D)$ is an operator built from these
two operators. A detailed introduction to this theory
is given in Appendix~\ref{app:kn}.

This theory provides a rigorous mathematical framework
for computing the plasma current.
In spectral full-wave codes, the real-space plasma
current is obtained by taking the product of the
conductivity-tensor spectrum at position $x$,
${\sigma}(x, k_x)$, with the electric-field spectrum,
${E}(k_x)$, and applying the inverse Fourier
transform~\cite{jaeger2001all}.
\begin{equation}
J(x)=\hat\sigma E(x) = \int_{-\infty}^{+\infty} e^{ikx}
\sigma(x, k) \hat{E}(k) \, dk
\label{eq:spectral-current}
\end{equation}
This is exactly the operator action defined by
Eq.~\eqref{eq:kn-quantization}.

Within this framework, the clearing of the denominator
that relates Eq.~\eqref{eq:fluid-unit} to Eq.~\eqref{eq:naive-clearing} cannot
be regarded as an algebraic multiplication of
$k_x - c_l$ on both sides, but is instead the
application of the operator $\widehat{k_x - c_l} = -i\partial_x - c_l(x)$
to both sides of the equation.
\begin{equation}
\widehat{(k_x - c_l)}\, \boldsymbol{v}_l =
\widehat{(k_x - c_l)}\,
\widehat{\left(\frac{\overset{\leftrightarrow}{b}_l}
{k_x - c_l}\right)} \cdot \boldsymbol{E}
\label{eq:operator-action}
\end{equation}

Cast into an ordinary differential equation, the
left-hand side of Eq.~\eqref{eq:operator-action} follows
directly from $k_x \rightarrow -i\partial_x$, whereas
its right-hand side involves the composition of the two
operators.
Under the present framework this composition is not
equivalent to the formal algebraic product: in addition
to the term obtained by ordinary multiplication, a
correction term arises from the ordering of the
operators\cite{folland1989harmonic}.
The detailed derivation is given in
Appendix~\ref{app:kn}. Including this correction term,
the corrected fluid response equation becomes
\begin{equation}
(-i\partial_x - c_l) \boldsymbol{v}_l =
\overset{\leftrightarrow}{b}_l \cdot \boldsymbol{E}
- i(\overset{\leftrightarrow}{b}_l' \cdot
\overset{\leftrightarrow}{b}_l^{-1}) \cdot
\boldsymbol{v}_l
\label{eq:corrected-response}
\end{equation}

\subsection{Numerical Implementation}

Equation~\eqref{eq:corrected-response} completes the 
plasma response. To obtain the wave field it must be 
solved together with the frequency-domain 
Maxwell equations:

\begin{equation}
\nabla \times (\nabla \times \boldsymbol{E}) -
\frac{\omega^2}{c^2} \boldsymbol{E}
= i \omega \mu_0 \sum_l \boldsymbol{v}_l
\label{eq:maxwell}
\end{equation}

With $\partial_y = 0$ and $\partial_z = i k_z$, the
operator $\nabla \times (\nabla \times \boldsymbol{E})$ in
Eq.~\eqref{eq:maxwell} retains the toroidal-curvature terms
$1/r$ and $1/r^2$ of the cylindrical geometry at
$r = R_0 + x$, its radial derivative entering only through
the combination $d^2/dx^2 + r^{-1} d/dx =
r^{-1} \frac{d}{dx}\left(r \frac{d}{dx}\right)$.

These coupled equations are discretized by finite 
differences on a uniform grid in x, with grid coordinates
$x_i = -a + (i-1) \Delta x$ ($i = 1, \dots, N$) and grid
spacing $\Delta x = 2a / (N-1)$. All
derivatives are approximated by second-order central
differences:

\begin{equation}
\begin{aligned}
\left. \frac{df}{dx} \right|_i &\simeq
\frac{f_{i+1} - f_{i-1}}{2\Delta x}, \\
\left. \frac{d^2 f}{dx^2} \right|_i &\simeq
\frac{f_{i-1} - 2f_i + f_{i+1}}{\Delta x^2}
\end{aligned}
\label{eq:finit-difference}
\end{equation}

Applying Eq.~\eqref{eq:finit-difference} to every interior grid point
of Eqs.~\eqref{eq:corrected-response} and \eqref{eq:maxwell},
together with the boundary conditions in 
Appendix~\ref{app:simu}, the equations are
assembled into a large sparse linear system, which
will be solved by the Intel MKL PARDISO solver:
\begin{equation}
\boldsymbol{A} \boldsymbol{X} = \boldsymbol{b}
\label{eq:linear-system}
\end{equation}

The solution vector $\boldsymbol{X}$ arranges all field
quantities in grid-point order:
\begin{equation*}
\begin{aligned}
\boldsymbol{X} = [&E_x^1, E_y^1, E_z^1, v_{lx}^1, v_{ly}^1,
v_{lz}^1, \\
&\dots, E_x^N, E_y^N, E_z^N, v_{lx}^N, v_{ly}^N,
v_{lz}^N]^T
\end{aligned}
\end{equation*}

where the superscript denotes the grid-point index. For
$S$ particle species, a Bessel harmonic
truncation $N_{\text{bessel}}$, and expansion orders
$l_n$ for each $\Gamma_n$, the total number of
field components and current expansion components at a
single grid point is $3 + 3 N_l$, where $N_l
= \sum_s \sum_{n=0}^{N_{\text{bessel}}} l_n$. The total
matrix dimension is $N \times (3 + 3 N_l)$.
Under typical JET D--H parameters, $N$ is of the order
of several thousand and $N_l \approx 200$,
giving a total dimension of roughly $10^6$--$10^7$.

This completes the full-wave solver BO-FW1D, 
which yields wave fields that retain the
key kinetic effects, including fast-wave propagation,
mode conversion, and the propagation and damping of
the IBW. These results will be presented and
benchmarked in detail in
the next section against the all-orders
spectral benchmark code~\cite{bude2021accelerating}.
The simulation geometry and plasma profiles are given
in Appendix~\ref{app:simu}.

\section{Results}
\label{sec:results}

\subsection{Validation against the All-Orders Kinetic
Benchmark}
\label{sec:validation}

\begin{figure}[t]
  \centering
  \includegraphics[width=\textwidth]{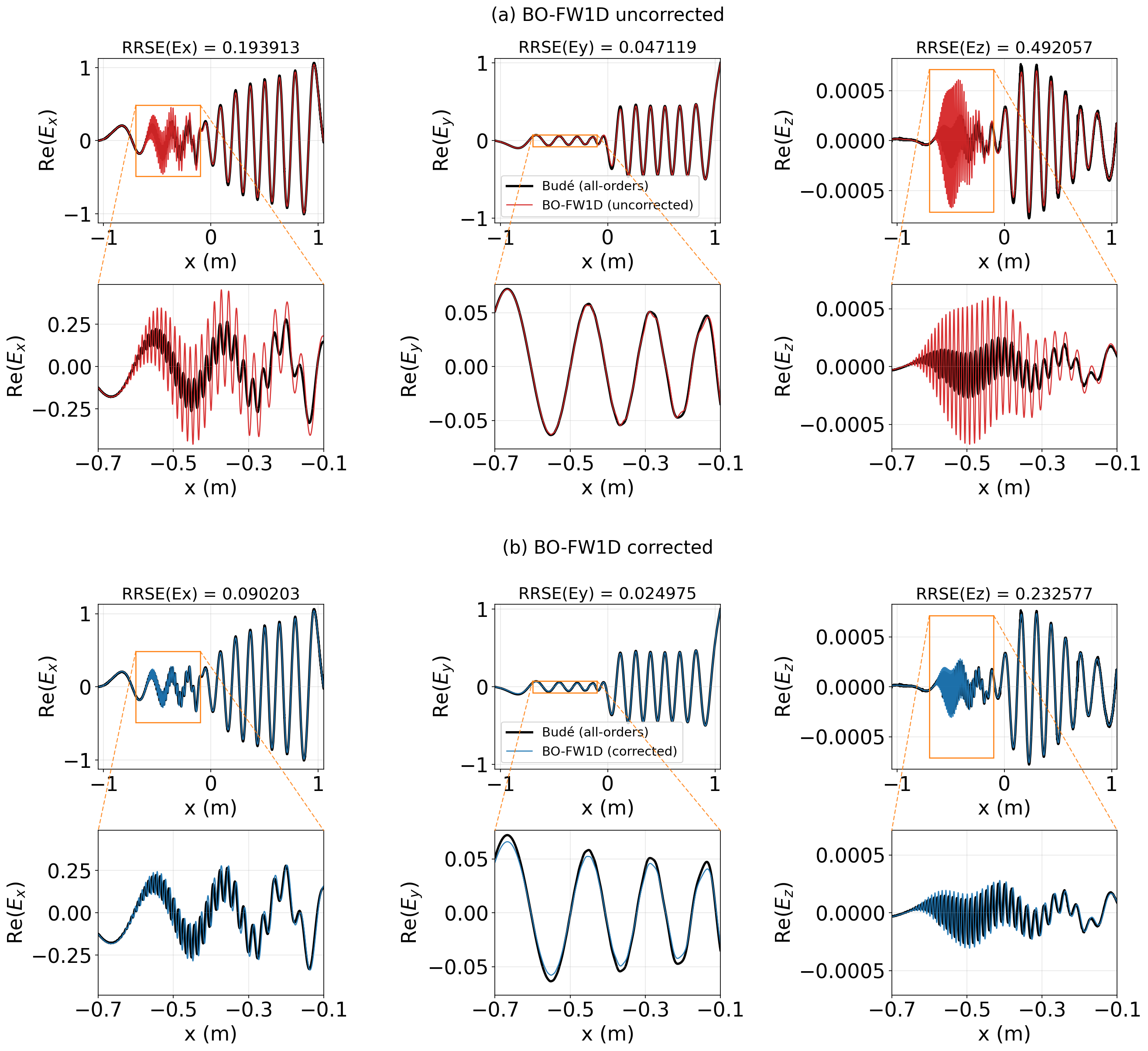}
  \caption{Real parts of the electric-field components
  of the JET D--H case of Table~\ref{tab:jet-parameters}: (a) BO-FW1D
  without and (b) BO-FW1D with the operator-composition
  correction term. In each block the upper
  row shows $E_x$, $E_y$ and $E_z$ over the full
  simulation domain, RRSE with the benchmark are quoted 
  in the title. The lower row enlarges the
  short-wavelength IBW region 
  $-0.7\,\text{m} \le x \le -0.1\,\text{m}$.}
  \label{fig:field-profiles}
\end{figure}

BO-FW1D is
compared directly with the spectral full-wave code of
Ref.~\cite{bude2021accelerating} (all-orders kinetic
equations, spectral method), the two codes being run
independently under the JET D--H parameters of
Table~\ref{tab:jet-parameters}, with BO-FW1D using $N = 5000$ grid
points. We apply the Root Relative Squared Error (RRSE)
to quantitatively measure the difference between the two
results:
\begin{equation*}
\text{RRSE}_\alpha = \sqrt{
\frac{\sum_i |E_\alpha^{\text{sim}}(x_i) -
E_\alpha^{\text{ref}}(x_i)|^2}
{\sum_i |E_\alpha^{\text{ref}}(x_i) -
\overline{E_\alpha^{\text{ref}}}|^2}
}, \quad \alpha = x, y, z
\end{equation*}

where $\overline{E_\alpha^{\text{ref}}}$ is the mean of
the reference electric field over the
simulation domain. Following
Ref.~\cite{bude2021accelerating}, the sums run over the
core plasma only, $|x| < a_p = 0.95$~m, and the RRSE is
evaluated separately for each component, since their
amplitudes differ by orders of magnitude.

From Figure~\ref{fig:field-profiles}, the wave amplitudes and phases
of the two codes nearly coincide in the
fast-wave propagation region ($x > 0\,\text{m}$).
However, in the IBW region on the $x < 0$ side, the
amplitude of the uncorrected BO-FW1D is clearly higher
than the all-orders baseline. After the
correction term is introduced, the BO-FW1D result
recovers to nearly coincide with the benchmark: near the
magnetic axis both consistently capture the mode
conversion from the fast wave to the IBW, the
short-wavelength IBW oscillations are clearly
distinguishable on the $x < 0$ side and gradually
decaying with propagation, and the amplitude and decay
envelope of the IBW are identical.

Quantitatively measuring the correction effect, the
RRSEs of all three electric-field components decrease
with respect to the all-orders kinetic benchmark, among
which the RRSE of $E_x$ drops from $0.194$ to $0.090$
(about $54\%$). This result fully
demonstrates that the operator composition correction
term is necessary for the finite-difference
code to obtain correct results, and the
mathematical treatment of the equations is
self-consistent and reliable when solving the fluid BW
dielectric tensor.

\subsection{Computational Scaling and Grid Convergence}
\label{sec:scaling}

The comparisons above were all performed at a single
grid resolution, so what remains to be established is
whether the finite-difference formulation is convergent
under grid refinement and whether the cost of the solve
is indeed only linear in the number of grid points $N$,
as anticipated from the sparse banded structure of
Eq.~\eqref{eq:linear-system}. To this end, the JET D--H case of
Table~\ref{tab:jet-parameters} is repeated from $N = 1500$ to $7000$ in
steps of $\Delta N = 100$, with all
remaining numerical settings and the boundary treatment
kept unchanged. The reference solution is the all-orders
kinetic spectrum of Ref.~\cite{bude2021accelerating},
reconstructed separately on the grid of each run so that
every $N$ is compared at its own resolution.

\begin{table}[H]
  \centering
  \caption{Computational cost of the two codes.}
  \label{tab:cost}
  \begin{tabular}{@{}lcc@{}}
    \toprule
    & BO-FW1D & All-orders \\
    \midrule
    Total cost         & $\propto N^{0.97}$ & $\propto N^{2.09}$ \\
    Matrix solve       & $\propto N^{0.98}$ & $\propto N^{3.01}$ \\
    Matrix structure   & banded & dense \\
    $N = 1500$         & $2.58\,\mathrm{s}$ & $8.96\,\mathrm{s}$ \\
    $N = 7000$         & $11.2\,\mathrm{s}$ & $157\,\mathrm{s}$ \\
    \bottomrule
  \end{tabular}
\end{table}

\begin{figure}[t]
  \centering
  \begin{subfigure}[t]{0.32\textwidth}
    \centering
    \includegraphics[width=\linewidth]{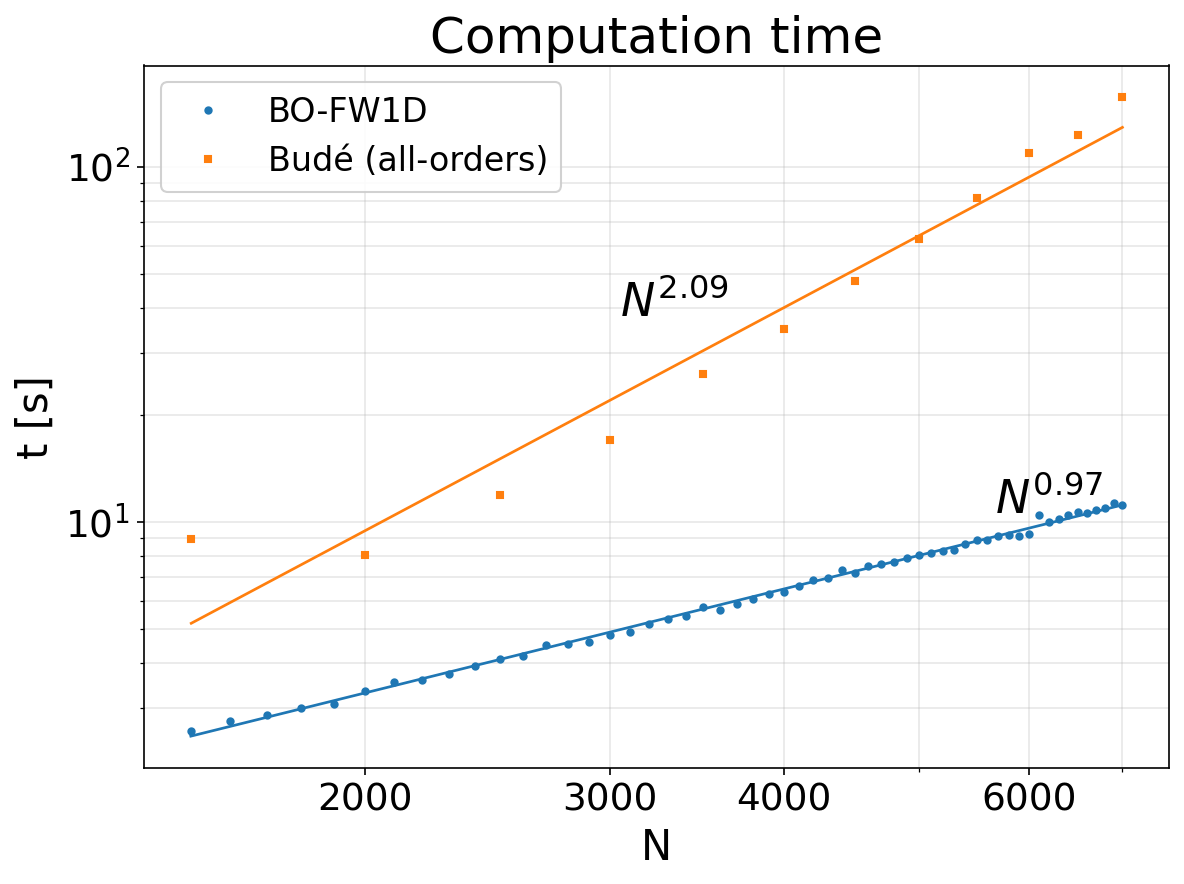}
    \caption{Total computation time}
    \label{fig:scaling-time}
  \end{subfigure}\hfill
  \begin{subfigure}[t]{0.32\textwidth}
    \centering
    \includegraphics[width=\linewidth]{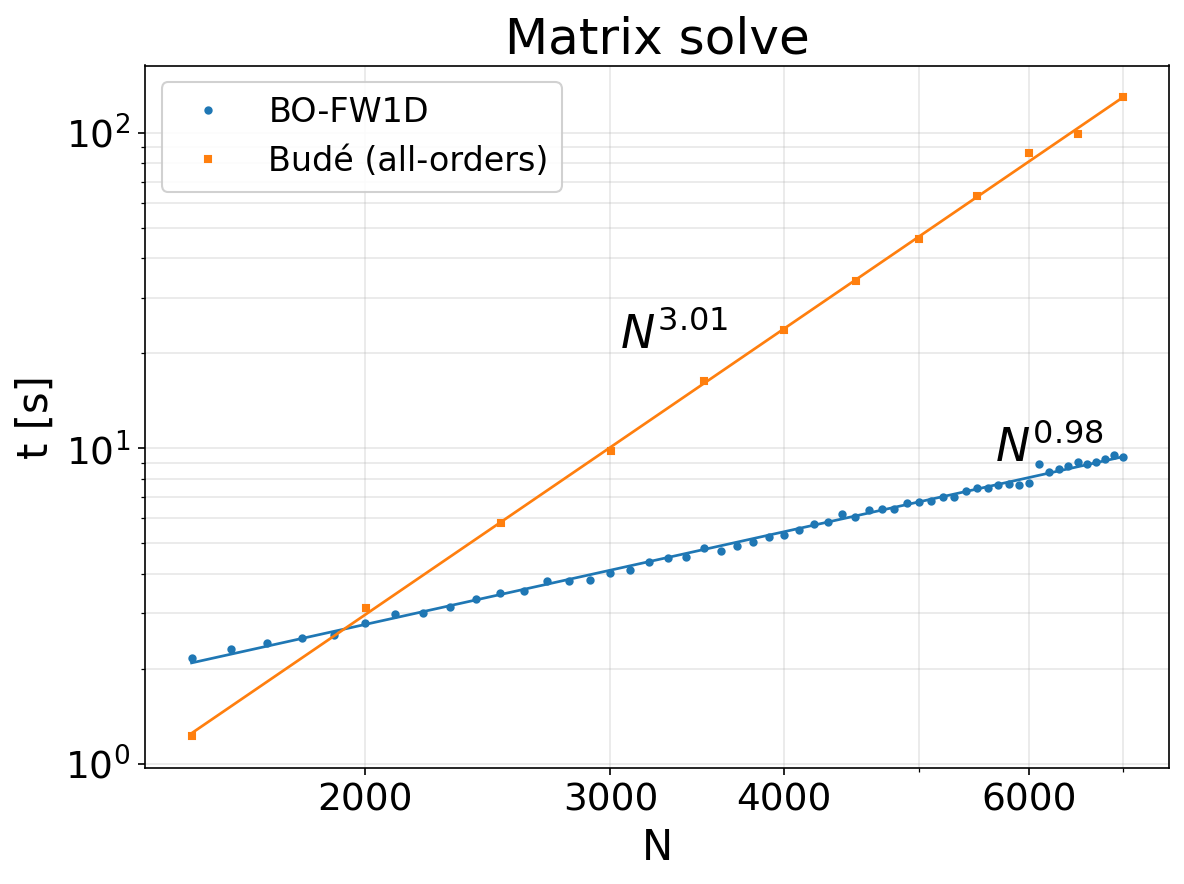}
    \caption{Matrix solve time}
    \label{fig:scaling-solve}
  \end{subfigure}\hfill
  \begin{subfigure}[t]{0.32\textwidth}
    \centering
    \includegraphics[width=\linewidth]{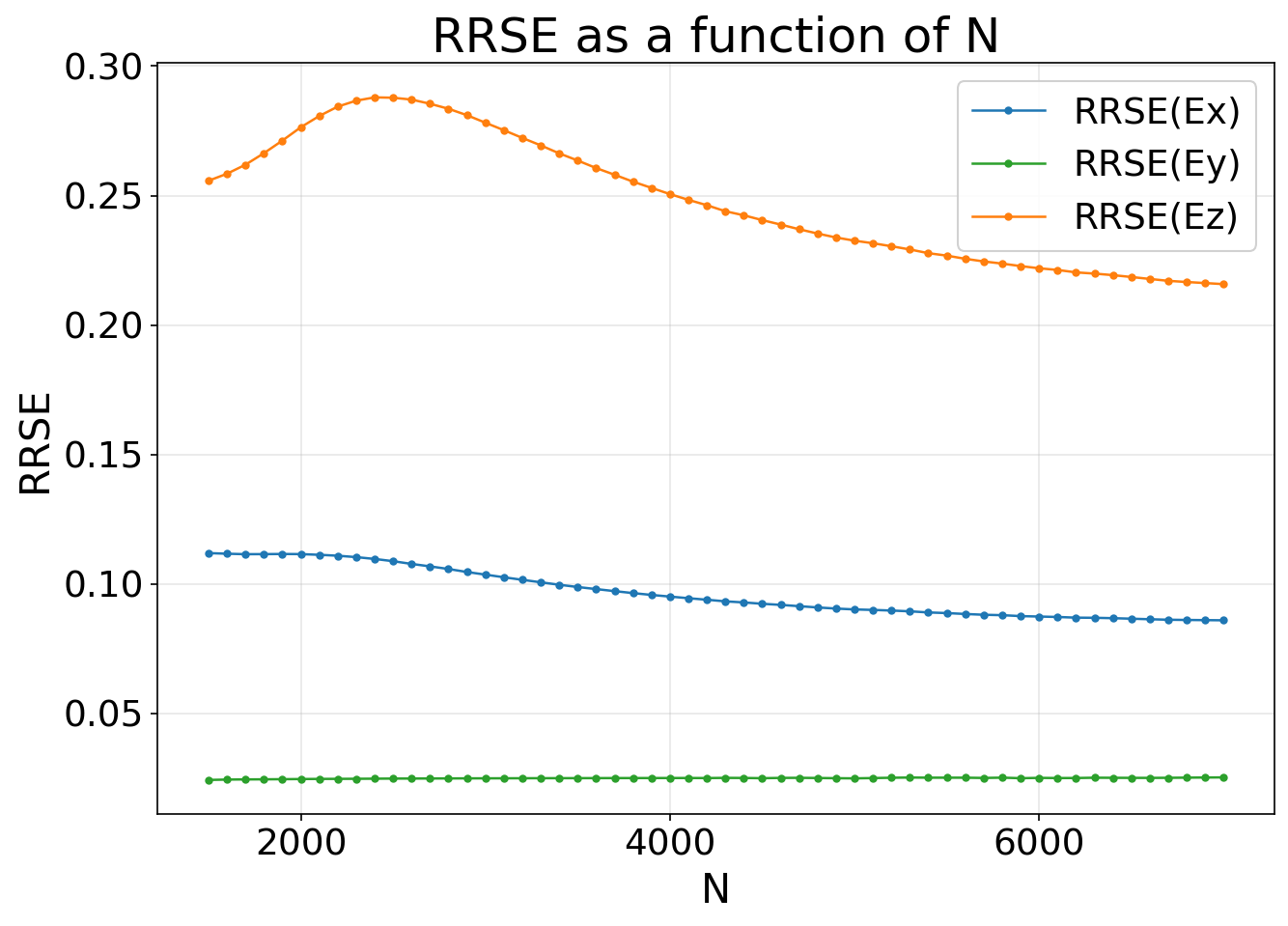}
    \caption{RRSE against the kinetic reference}
    \label{fig:scaling-rrse}
  \end{subfigure}
  \caption{Grid-refinement study of BO-FW1D under the
  JET D--H parameters: (a) total
  computation time (b) matrix solve time as
  functions of the number of grid points $N$, together
  with the power-law fit $aN^{b}$ (solid curve)
  (c) RRSE of the three electric-field components
  with respect to the all-orders reference.}

  \vspace{1ex}
  \begin{minipage}{\textwidth}
    \setlength{\leftskip}{0pt}
    \setlength{\rightskip}{0pt plus 2em}
    \setlength{\parfillskip}{0pt plus 1fil}
    \small (Environment: a laptop with Intel
    i9-14900HX and 32~GB memory.
    All runs use identical compiler settings.)
  \end{minipage}
  \label{fig:scaling}
\end{figure}

Figure~\ref{fig:scaling-time} compares the total computation time
of the two codes, and Figure~\ref{fig:scaling-solve} solely compares
the time for solving the matrix. The fitted power laws
$y = aN^{b}$ and typical computation times
are recorded in Table~\ref{tab:cost}.
These measurements show that BO-FW1D attains close to
linear scaling in $N$ both in the matrix solve and in the
total computation time. 
For the all-orders code, the total computational
cost approximately follows $O(N^{2})$, and the matrix
solving process follows the $O(N^{3})$ law reported in
Ref.~\cite{bude2021accelerating}. Consequently,
the speed advantage of BO-FW1D over the all-orders code
grows with the number of grid points.

Figure~\ref{fig:scaling-rrse} turns to the accuracy side of the
same scan. Since the IBW is, to leading order, a
quasi-electrostatic mode that
propagates nearly perpendicular to the magnetic field
with $k_\perp\rho_i \gtrsim 1$~\cite{brambilla1999numerical},
its electric field is dominated by the component along
$\mathbf{k}_\perp$; in the present slab geometry
($\mathbf{B} = B\hat{z}$, $k_y = 0$) this is $E_x$, so
that the $\mathrm{RRSE}$ of $E_x$ is the quantity of interest.
It decreases with $N$ overall, from $0.112$ at
$N = 1500$ to $0.086$ at $N = 7000$, a reduction of about
$23\%$ over the window. The two remaining components are
of secondary interest only. $E_z$ has an amplitude far
smaller than those of $E_x$ and $E_y$, since the very
large conductivity parallel to the magnetic field
strongly suppresses the parallel electric field, so that
its relative error has little bearing on the accuracy of
the wave field; moreover, for the present JET case $E_y$
does not carry the mode-converted
IBW~\cite{bude2021accelerating}. Consistent with this,
$\mathrm{RRSE}(E_y)$ is essentially independent of $N$
and stays close to $0.025$ throughout the whole window.
The $N$-dependence of the error is therefore confined to a
small residual decrease, and the solution is already close
to convergence.

Taken together, these results show that the fluid BW 
formulation retains the full kinetic response of the 
plasma--the mode conversion, the short-wavelength IBW 
and its damping--while requiring only 
a sparse banded solve whose cost grows almost linearly 
with the number of grid points. Calculating full kinetic 
wave physics therefore no longer demands a dense, 
globally coupled response matrix.

\section{Conclusion and Discussion}
\label{sec:conclusion}

Building on the fluid BW dielectric tensor proposed by
Xie~\cite{xie2025developing}, this work develops a
novel way of including kinetic effects in full-wave
simulations with a fluid model. It establishes, and
validates, a computational paradigm that extends the
local kinetic dispersion relation to the full-wave
solution in inhomogeneous media in the form of real-space
differential equations, develops the one-dimensional
frequency-domain finite-difference code BO-FW1D,
and carries out a systematic validation under JET D--H
plasma parameters.

The main findings and conclusions of this work are as
follows. The local dispersion relation of the fluid BW
model cannot be
straightforwardly extended to differential equations by
algebraic manipulation; if so, the non-commutativity of
the multiplication order would give rise to a
non-physical amplification of the IBW amplitude. The
pseudodifferential operator theory based on
Kohn-Nirenberg ordering offers a systematic treatment of
this problem, and the operator-composition correction
term restores the finite-difference results to agreement
with the all-orders benchmark. In the
resulting code, full-wave simulation using the fluid BW
dielectric tensor is able to capture all key kinetic
effects, with central differencing yielding a sparse
banded matrix system whose solution cost grows only
linearly with $N$ ($b \simeq 0.97$), and grid refinement
driving the solution towards convergence.

There is still much to be done. As the matrix structure
is fixed, tailoring the solver according
to the fixed structure of the matrix would bring
a substantial speed-up. Meanwhile, a systematic 
scan of the
L-pole expansion order L and truncated Bessel
harmonic order N would quantify the accuracy--cost
trade-off. Moreover, the fluid BW model's 
application in other frequency regimes is also
worth exploring. The simulation results are
currently restricted to one-dimensional
slab geometry, the $k_y = 0$ assumption, and
a simplified antenna model. Because the speed 
advantage over the all-orders code grows with N, 
extending the formulation to the more grid-intensive 
two-dimensional toroidal geometry 
is a particularly promising next step.

\appendix

\makeatletter
\renewcommand{\@seccntformat}[1]{%
  \expandafter\ifx\csname the#1\endcsname\thesection
    \raggedright\hyphenpenalty=10000\exhyphenpenalty=10000
    Appendix~\csname the#1\endcsname\quad
  \else
    \csname the#1\endcsname\quad
  \fi}
\def\@hangfrom#1{\noindent #1}
\makeatother

\section{Pseudodifferential Operator Correction Based
on Kohn-Nirenberg Ordering}
\label{app:kn}

The core of pseudodifferential operator theory is the
quantization problem: associating to each
function $\sigma(x, k)$ on phase space an operator
$\sigma(X, D)$ acting on the function space, such
that the coordinate function $x$ corresponds to the
coordinate operator $X$ and the wavenumber
function $k$ corresponds to the differential operator $D
= -i\partial_x$. Conversely, the
phase-space function $\sigma(x, k)$ associated with a
given operator is called its symbol. When the
symbol is a polynomial in $k$, the corresponding
operator reduces to an ordinary differential
operator; when the symbol is not a polynomial---e.g., a
rational fraction---the corresponding
operator is called a pseudodifferential operator. Since
$X$ and $D$ do not commute, the operator
associated with a given symbol depends on the order in
which $x$ and $k$ act. The
Kohn-Nirenberg ordering places the coordinate
operator before the differential operator, so that the
$k$-dependence of the symbol acts on the
field before its $x$-dependence.

Pseudodifferential operator theory provides a rigorous
mathematical framework for handling the
real-space counterparts of non-polynomial functions on
phase space. Under the KN
ordering, the real-space operator corresponding to a
phase-space function $\sigma(x, k)$ is the one defined
by Eq.~\eqref{eq:kn-quantization}: the symbol is convolved with the
spectrum of $f(x)$, its $k$-dependence being applied
before its $x$-dependence.

As mentioned in the main text, the so-called
denominator-clearing process introduces a composite
operator

\begin{equation}
\mathcal{P} = (k_x - c_l) \circ \left(
\frac{\overset{\leftrightarrow}{b}_l}{k_x - c_l} \right)
\label{eq:composite-op}
\end{equation}

In the pseudodifferential operator framework, the
composition of two operators follows
(Ref.~\cite{folland1989harmonic},Theorem 2.52):

\begin{equation}
\begin{aligned}
\sigma_{A \circ B} ={} & \sigma_A \cdot \sigma_B
- i \frac{\partial \sigma_A}{\partial k}
\frac{\partial \sigma_B}{\partial x} \\
& + \frac{(-i)^2}{2!} \frac{\partial^2 \sigma_A}{\partial k^2}
\frac{\partial^2 \sigma_B}{\partial x^2}
+ \cdots
\end{aligned}
\label{eq:composition-rule}
\end{equation}

In inhomogeneous media, $b_l$ and $c_l$ are functions
of x, so the correction terms cannot be
neglected. Since $\sigma_A = k_x - c_l(x)$ is merely
linear in $k$, $\partial^2 \sigma_A / \partial
k^2 = 0$, hence all correction terms of second order and
higher in Eq.~\eqref{eq:composition-rule} vanish
identically, and the first-order correction already
gives the exact composition result:

\begin{equation}
\begin{aligned}
\delta\mathcal{P} &= -i \cdot \frac{\partial}{\partial k}(k_x - c_l)
\cdot \frac{\partial}{\partial x}\!\left(
\frac{\overset{\leftrightarrow}{b}_l}{k_x - c_l} \right) \\
&= -i \frac{\overset{\leftrightarrow}{b}_l'}{k_x - c_l}
- i \frac{\overset{\leftrightarrow}{b}_l \, c_l'}{(k_x - c_l)^2}
\end{aligned}
\label{eq:delta-P}
\end{equation}

Here, $\overset{\leftrightarrow}{b}_l' \equiv \partial
\overset{\leftrightarrow}{b}_l / \partial x$
and $c_l' \equiv \partial c_l / \partial x$. The second
term in Eq.~\eqref{eq:delta-P} contains $c_l'$;
since $c_l$ depends only on the Larmor radius
$\rho_{cs}$ and thus varies slowly in space, this term
is a small quantity under typical tokamak parameters and
can be neglected. Using $\boldsymbol{v}_l =
[\overset{\leftrightarrow}{b}_l / (k_x - c_l)] \cdot
\boldsymbol{E}$, the first term can be written
as Eq.~\eqref{eq:correction-term}, which is the correction term
in Eq.~\eqref{eq:corrected-response} of the main text:
\begin{equation}
\delta\mathcal{P} \, \boldsymbol{E} =
-i(\overset{\leftrightarrow}{b}_l' \cdot
\overset{\leftrightarrow}{b}_l^{-1}) \cdot
\boldsymbol{v}_l
\label{eq:correction-term}
\end{equation}

\section{Simulation Geometry and Plasma Profiles}
\label{app:simu}

To facilitate benchmarking against existing full-wave
codes, the simulation geometry and plasma
profiles are chosen to be identical to those of
Ref.~\cite{bude2021accelerating}, i.e., typical
parameters of the JET tokamak. Taking the magnetic axis
as the radial coordinate origin ($x = R - R_0$), the
major radius is $R_0 = 2.97\,\text{m}$, the plasma
minor radius is $a = 1.05\,\text{m}$, and the
simulation domain is $x \in [-a, a]$.

The magnetic field contains only the toroidal component,
decaying radially as $1/R$:
\begin{equation*}
B_z(x) = B_0 \frac{R_0}{R_0 + x}, \quad B_0 =
3.45\,\text{T}
\end{equation*}

The ICRH antenna operates at frequency $f =
51\,\text{MHz}$ with toroidal mode number
$n_{\text{tor}} = 27$, corresponding to the JET antenna
dipole phasing mode. The parallel wavenumber
varies radially as $k_z(x) = n_{\text{tor}} / (R_0 + x)$.

The plasma consists of fully ionized deuterium ions
(95\%) and a hydrogen minority (5\%), with the electron
density determined by the quasi-neutrality condition
$n_e = n_D + n_H$. Both the density and
temperature profiles adopt a piecewise form with a
parabolic-like core and an exponentially decaying edge:
\begin{equation*}
\begin{aligned}
G_{\text{core}}(x) &= (G_0 - G_s)\left(1 - (x /
a_p)^2\right)^{\alpha} + G_s \\
G_{\text{edge}}(x) &= G_s \exp\!\left(-\frac{|x -
a_p|}{\lambda}\right)
\end{aligned}
\end{equation*}

In the above, $G_0$ is the value at the magnetic axis,
$a_p$ is the core--edge transition point,
$G_s$ is the profile value at that point, and $\alpha$
and $\lambda$ control the shapes of the core
and edge regions, respectively. The specific parameters
are listed in Table~\ref{tab:jet-parameters}.
\begin{table}[H]
  \centering
  \caption{Simulation parameters of the JET D--H plasma}
  \label{tab:jet-parameters}
  \begin{tabular}{@{}ccc@{}}
    \toprule
    Symbol & Density & Temperature \\
    \midrule
    $G_0$ & $n_0 = 7\times10^{19}\,\mathrm{m^{-3}}$
    & $T_0 = 5\,\mathrm{keV}$ \\
    $G_s$ & $n_s = 2\times10^{19}\,\mathrm{m^{-3}}$
    & $T_s = 100\,\mathrm{eV}$ \\
    $\alpha$ & $\alpha_n = 1$ & $\alpha_T = 1.5$ \\
    $\lambda$ & $\lambda_n = 0.05\,\mathrm{m}$
    & $\lambda_T = 0.05\,\mathrm{m}$ \\
    $a_p$ & $a_p = 0.95\,\mathrm{m}$
    & $a_p = 0.95\,\mathrm{m}$ \\
    \bottomrule
  \end{tabular}
\end{table}

The electron and ion temperatures use the same profile
parameters. The density profiles of the
deuterium and hydrogen ions are assigned as $n_D(x) =
0.95\,n(x)$ and $n_H(x) = 0.05\,n(x)$.

The boundary conditions are set as follows. At $x = \pm
a$, $E_y$, $E_z$, and all current expansion
components $\boldsymbol{v}_l$ satisfy homogeneous
Dirichlet conditions; the $x$-derivative of $E_x$
is zero, implemented via a first-order one-sided
difference: $E_x^1 = E_x^0$ and $E_x^{N-1} =
E_x^N$, where $x_0$ and $x_{N+1}$ lie outside the
solution domain and serve only to close the
boundary difference stencils. The antenna excitation is
introduced by modifying the right-hand-side
vector $\boldsymbol{b}$: the right-hand side of the
equation corresponding to $E_y^N$ on the right
boundary is set to $1$, so that a tangential electric
field of unit amplitude models the radio-frequency
excitation of the antenna at the plasma edge.

\ack{The authors would like to thank the discussions
and comments from Haojie Ma, Yilin Li, Yuxuan Wang,
Jian Bao, Tieshuan Fan, Andong Xu, Beining Chen
and Ziheng Zhou.}

\funding{This work was supported by the National MCF Energy
R\&D Program of China (Grant No. 2025YFE03000100)}

\renewcommand{\bibsection}{\section*{References}}

\bibliographystyle{unsrtnat}
\bibliography{ref}

\end{document}